\documentclass{IEEEoj}

\usepackage{amsmath,amssymb,amsfonts}
\usepackage{algorithmic}
\usepackage{graphicx}
\usepackage{textcomp}
\usepackage{stfloats}
\usepackage{placeins}
\usepackage{float}
\usepackage{multirow}
\usepackage{glossaries}
\newacronym{ied}{IED}{Intelligent Electronic Device}
\newacronym{vied}{vIED}{Virtual IED}
\newacronym{sv}{SV}{Sampled Values}
\newacronym{goose}{GOOSE}{Generic Object-Oriented Substation Event}
\newacronym{kvm}{KVM}{Kernel-Based Virtual Machine}
\newacronym{vm}{VM}{Virtual Machine}

\usepackage{cite}
\usepackage{amsmath,amssymb,amsfonts}
\usepackage{algorithmic}
\usepackage{graphicx,color}
\usepackage{textcomp}
\usepackage{hyperref}
\def\BibTeX{{\rm B\kern-.05em{\sc i\kern-.025em b}\kern-.08em
    T\kern-.1667em\lower.7ex\hbox{E}\kern-.125emX}}
\AtBeginDocument{\definecolor{green}{cmyk}{1,0,1,0.5}}
\def\OJlogo{\vspace{-14pt}\includegraphics[height=28pt]{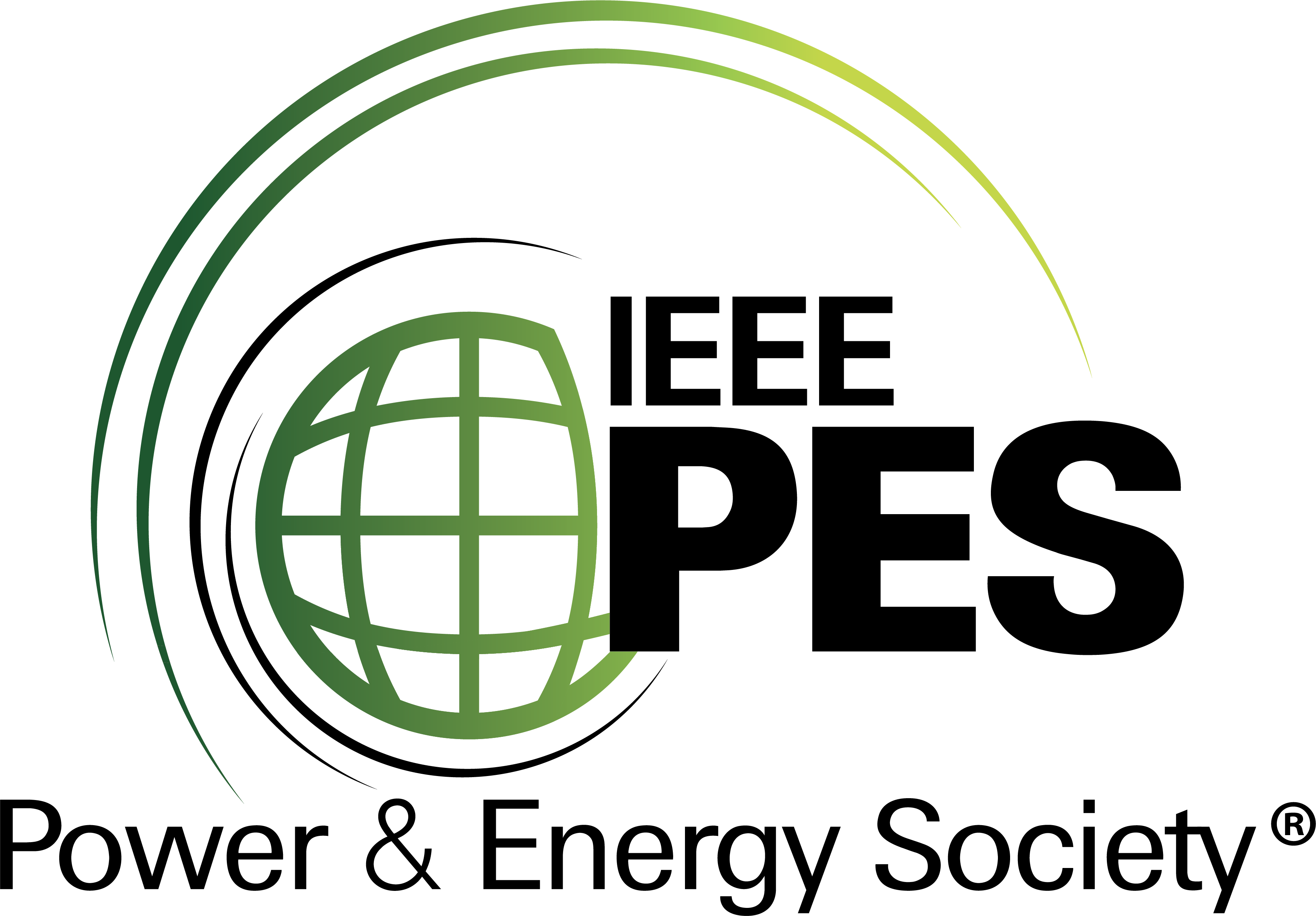}}
\begin{document}
%\receiveddate{XX Month, XXXX}
%\reviseddate{XX Month, XXXX}
%\accepteddate{XX Month, XXXX}
%\publisheddate{XX Month, XXXX}
%\currentdate{XX Month, XXXX}
%\doiinfo{OAJPE.2023.1234567}

\title{Design and Performance Assessment of a Virtualized IED for Digital Substations}

\author{\uppercase{Alailton J. Alves Junior}\authorrefmark{1},
\uppercase{Denis V. Coury}\authorrefmark{1}, MEMBER, IEEE, and \uppercase{Ricardo A. S. Fernandes}\authorrefmark{1}, SENIOR MEMBER, IEEE}

% \address[1]{Department of Electrical and Computer Engineering, São Carlos School of Engineering (EESC), University of São Paulo (USP), São Carlos, SP 13566-590, Brazil}
\corresp{Corresponding author: Alailton J. Alves Júnior (e-mail: alailtonjunior@usp.br)}
\affil{Department of Electrical and Computer Engineering, São Carlos School of Engineering (EESC), University of São Paulo (USP), São Carlos, SP 13566-590, Brazil}
\authornote{This work was supported by National Council for Scientific and Technological Development (CNPq) [402334/2023-0]. The authors would also like to thank the São Carlos School of Engineering, University of São Paulo, São Carlos, Brazil for the facilities provided.}
\markboth{Preparation of Papers for IEEE PES JOURNALS}{Author \textit{et al.}}

\begin{abstract}
Digital substations have significantly enhanced power grid protection by replacing traditional copper wiring with fiber-optic communication and integrating IEC 61850-compliant Intelligent Electronic Devices (IEDs), resulting in greater efficiency, reliability, and interoperability. While these advancements provide improved interoperability, challenges such as high costs, complex networks, and limited upgradeability persist. To mitigate these issues, the virtualization of IEDs has emerged as a cost-effective solution, offering scalability, simplified maintenance, and reduced hardware costs by replacing traditional hardware-based IEDs with software-based counterparts. However, the performance and reliability of virtual IEDs (vIED) must be rigorously evaluated to ensure their robustness in real-time applications. This paper develops, implements, and evaluates a vIED designed to match the performance of its hardware-based counterparts. The vIED was deployed on a server using virtual machines, with its core logic implemented in low-level programming languages to ensure high-speed, deterministic behavior. The performance was evaluated using real-time simulations, focusing on the response times of the protection functions. The results demonstrated that vIEDs achieved acceptable response times, validating their suitability for deployment in critical time-sensitive environments within digital substations.
\end{abstract}



\maketitle

\section{Introduction}
\IEEEPARstart{T}{he} electricity industry is undergoing a significant transformation driven by widespread digitalization, which is turning substations into smarter and more integral components of the modern power grid~\cite{Matanov2022}. This shift aims to improve efficiency, reliability, and security by replacing traditional copper-wired connections and conventional secondary equipment with fiber optic communication and \glspl{ied}. Operating under the IEC 61850 standard, these \glspl{ied} are a critical enabler of this transition, facilitating the integration of protection, control, and monitoring functions within a unified network~\cite{Santos2024}.

Although digital substations offer advantages such as reduced wiring, interoperability, and improved overall performance, the transition to these systems poses significant challenges. The high cost of \glspl{ied}, coupled with complex communication networks, can increase maintenance expenses and require specialized engineering expertise. In addition, integrating new technologies into existing infrastructure can be complex and costly. System upgrades are often complicated and can impact the equipment's lifecycle. Moreover, limitations in upgradeability for many digital devices can extend development cycles and inflate long-term maintenance costs~\cite{CIGREB560}.

To address these challenges, researchers have proposed the virtualization of \glspl{ied} as a cost-effective and flexible solution for digital substations. A software-based \gls{ied} offers several benefits, such as reduced hardware costs, simplified maintenance, and enhanced scalability~\cite{Guibout2024}. By emulating the functionality of physical \glspl{ied} in a software environment, \glspl{vied} can provide a more agile and adaptable protection system for power grids.

The remainder of this paper is organized as follows. Section~\ref{sec:related_works} presents the literature related to IEDs' virtualization. Section~\ref{sec:comparative_analysis} highlights the differences between physical and \glspl{vied}. Section~\ref{sec:architecture} describes the architecture of the \gls{vied}, detailing the protection algorithms and communication protocols implemented. Section~\ref{sec:test} presents the test system setup, including the simulated environment, hardware configuration, and network parameters. Next, Section~\ref{sec:performance} discusses the results, focusing on protection response times and communication delays. Finally, Section~\ref{sec:conclusions} draws conclusions, providing a concise summary of the main findings.

\section{Related Works}\label{sec:related_works}

A prominent area of research in power system protection focuses on using digital twins (DT) to enhance existing protection schemes. A DT is defined as a virtual representation of a physical asset or system, continuously updated with real-time data, allowing for comprehensive simulation, analysis, and optimization of the asset's performance throughout its lifecycle. These capabilities enable DTs to provide real-time simulation, scenario testing, and cost-effective validation of protection schemes, ultimately contributing to improved system reliability, reduced research costs, and more effective decision making~\cite{Jiang2022}. 

Recent research has also explored the potential of DTs in modernizing power grids. For example, in~\cite{Sifat2024}, the authors proposed a framework for Digital Twin Electric Grids, addressing key aspects such as subsystem-level modeling, modular design, and scalability. This framework integrates advanced real-time data processing with predictive maintenance and machine learning-based optimization techniques, ensuring enhanced grid stability and performance.  

\textcolor{black}{Concerning protection applications, the authors of~\cite{Gmez-Luna2023, Gmez-Luna2024} investigated using DT in overcurrent protection schemes for distribution networks integrated with distributed energy resources (DER)}. Their work introduced a validation approach for directional overcurrent protection in meshed distribution networks~\cite{Gmez-Luna2024} and proposed a DT-based overcurrent protection scheme for DER-integrated distribution networks~\cite{Gmez-Luna2023}. Both studies emphasized the role of DTs in improving fault isolation, reducing restoration times, and enabling cost-effective validation of protection schemes. Furthermore, a DT-based relay protection mirror operation technology was proposed in~\cite{Yao2023}, enhancing fault detection and system reliability through real-time monitoring and simulation. These studies emphasize the transformative impact of DTs on power grid protection and management.

Despite their ability to simulate real-time operations and predict failure modes, DT differ fundamentally from fully virtualized systems. Although they complement physical assets, they do not replace them. Thus, DT serves as a preliminary step towards achieving full system virtualization by providing a virtual model that interacts with real-world data, thus ensuring that the transition to complete virtualization or remote monitoring is both feasible and effective. However, it is important to note that the literature on fully virtualized IEDs remains scarce. Most existing research focuses on using DT as complementary tools, rather than as complete replacements for physical IEDs.

In contrast to DT, \glspl{vied} represent a more advanced approach, in which the functionality of physical IEDs is fully replicated in a virtual environment. Several studies have investigated the deployment of \glspl{vied} in substation protection systems, highlighting both their potential and the technical challenges associated with their implementation. Notably, the studies proposed in~\cite{Wojtowicz2018} and~\cite{Wojtowicz2022} examined the feasibility of deploying \gls{vied}, addressing critical factors such as time synchronization, communication delays, and \gls{goose} message transmission. These studies, which used virtualization platforms such as KVM (Kernel Virtual Machine), highlighted key areas for improvement, particularly in response time and synchronization accuracy.

Further comparative studies, such as those proposed by~\cite{Ansari2020} and~\cite{Rsch2024}, assessed the performance of \glspl{vied} versus that of physical IEDs in real-time simulations. Their findings indicated that \glspl{vied} can achieve comparable protection performance, including tripping times of the circuit breaker and communication. However, these studies also identified the need for continued advancements in time synchronization and modeling of more complex network and protection scenarios.

\textcolor{black}{A broader perspective on the underlying technology was offered in~\cite{Queiroz2024}, which presented a systematic review of container-based virtualization for real-time industrial systems. Their work summarizes techniques to ensure real-time compliance in virtual environments, which is fundamental for \glspl{vied}. The review identifies several key methods (including single-kernel methods using patches, co-kernel systems for handling time-critical tasks, and advanced scheduling-based methods). The study concludes that while containers offer near-native performance, the container-induced overhead delay can range from a few microseconds to over 100~$\mu$s depending on the configuration. The review also highlights open challenges in orchestration, deterministic inter-container communication, and security, which are directly relevant to the robust deployment of \glspl{vied}.}

Additionally, the deployment of virtualized protection systems in edge computing environments has been explored in the context of low-latency requirements for IEC~61850 communications. The authors of~\cite{Carvalho2022} investigated the scalability and efficiency of \glspl{vied} in such environments, emphasizing their performance in orchestrated settings. The study also highlighted the need for future enhancements, including the integration of security and redundancy mechanisms, to ensure the reliability and robustness of the system in dynamic operational conditions.

Building on these developments, in~\cite{Kabbara2024}, the authors proposed a hybrid framework that combines \glspl{vied} with software-defined networking (SDN) to enhance scalability and adaptability in IEC 61850 digital substations. The study, which used real-time simulations of the IEEE 5-bus model, demonstrated the framework's capacity to manage network complexity and support low-latency communication, illustrating the potential of SDN in optimizing virtualized protection systems.

These studies demonstrate that virtualization of \glspl{ied} offers a promising avenue to enhance the flexibility and scalability of digital substations. However, challenges remain in ensuring time synchronization, managing communication delays, addressing cybersecurity vulnerabilities, and integrating virtualized systems into existing infrastructure. The ongoing development of \glspl{vied}, particularly in low-latency and real-time environments, highlights the potential for these systems to meet the evolving demands of modern power grids, provided that rigorous testing and validation processes are in place to ensure their reliability and performance.

Considering the above context, this paper presents the development and implementation of an \gls{vied} for digital substation applications. The proposed \gls{vied} was designed to operate similarly to a physical \gls{ied}, implementing protection schemes and communication protocols in accordance with the IEC 61850 standard. Its performance was evaluated through real-time simulations, focusing on protection response times, communication delays, and system reliability. {\color{black} Therefore, the main contributions of this paper are: (i) the development of an open-source \gls{vied} that provides a transparent and adaptable platform, enabling the integration of well-established protection algorithms into a fully virtualized IEC 61850-compliant framework; (ii) a performance evaluation of the \gls{vied} in a reproducible virtualized testbed, considering different fault scenarios; and (iii) a discussion on the practical challenges of \gls{vied} deployment, including cybersecurity, reliability, and redundancy, contextualizing the results within real-world constraints.}

\section{A Comparative Analysis of Physical and Virtual IEDs}\label{sec:comparative_analysis}

{\color{black}
A physical \gls{ied} is a dedicated, ruggedized hardware device specifically engineered to perform protection and control functions within substation environments. These devices are equipped with specialized processors and input/output (I/O) interfaces, enabling fast and deterministic execution of protection algorithms. Directly interfacing with conventional current and voltage transformers, physical IEDs process raw analog signals locally and initiate protective actions with minimal and predictable latency. Modern physical IEDs typically support the IEC 61850 standard, allowing them to publish and subscribe to high-speed \gls{goose} and \gls{sv} messages for real-time communication with other devices.

In contrast, a \gls{vied} employs a software-defined architecture that operates on general-purpose computing platforms, such as \glspl{vm} or containers. These virtualized devices are typically hosted on centralized servers within substation control rooms and interface with Merging Units (MUs) and non-conventional instrument transformers via IEC 61850 protocols. Although vIEDs use the same communication standards as their physical counterparts, their performance is heavily influenced by the underlying Information Technology (IT) and network infrastructure. Consequently, factors such as network latency, data consistency, and cybersecurity become crucial considerations in their deployment.

Virtualization introduces several notable advantages, including improved scalability, reduced hardware footprint, and more efficient lifecycle management. One of the key benefits is the ability to consolidate protection and control functions from multiple physical devices onto centralized servers, thereby decreasing the amount of dedicated hardware required~\cite{Vilaplana2024}. Unlike physical systems, where expansion often involves procuring and installing new hardware, vIEDs can be scaled by deploying additional software instances on existing infrastructure. This shift enables a new economic model, trading high operational expenses for a more capital-intensive but scalable and cost-efficient approach~\cite{CIGREB560}.

Lifecycle management is fundamentally transformed by virtualization. The decoupling of protection logic from physical hardware enables agile software development and deployment practices previously unattainable in substations. This separation allows for independent testing, streamlined updates, and significantly faster deployment cycles, mitigating the hardware-software compatibility issues that plague physical systems. Updates can be deployed remotely and seamlessly, borrowing strategies from modern IT like blue-green deployments or canary releases, where a new vIED version can be tested on live data in parallel with the old one before a full switchover, drastically reducing risk and service disruption~\cite{WorkingGroup2016}. Furthermore, this software-defined nature innovates redundancy. Instead of relying on static hardware backups, protection functions can be dynamically replicated and orchestrated across multiple servers, enabling sophisticated, automatic failover schemes that enhance overall system resilience.

Despite these advantages, transitioning to virtualized architectures introduces significant challenges, particularly concerning performance and reliability. Physical IEDs achieve deterministic performance through dedicated hardware and direct, hardwired connections. In contrast, vIEDs depend on real-time computing guarantees within a shared environment, where performance must be managed at both the host (hypervisor) and guest (VM/container) levels~\cite{Guibout2024}. To meet the stringent timing requirements of protection functions, generic resource allocation is insufficient. Advanced virtualization techniques are required, such as Central Processing Unit (CPU) pinning (or CPU affinity), which dedicates specific processor cores to a vIED to eliminate scheduling latency and jitter caused by other processes. Similarly, for network I/O, technologies like Single Root I/O Virtualization (SR-IOV) can be employed to grant a vIED direct, low-latency access to the physical Network Interface Card (NIC), bypassing the hypervisor's virtual switch. The overall reliability of the system therefore depends on both the robustness of the computing infrastructure and the efficiency of the communication network.

Security considerations are significantly amplified in a virtualized setup. While physical IEDs benefit from the inherent security of physical isolation, vIEDs introduce new software layers—such as the hypervisor, host operating system, and orchestration platforms, that substantially expand the system's attack surface. The centralization of multiple vIEDs onto a single server, while efficient, creates a critical single point of failure. A successful intrusion into the virtualization host could allow an adversary to simultaneously compromise, disable, or manipulate multiple protection functions, potentially leading to a widespread outage or catastrophic equipment damage~\cite{Tobar-Rosero2025}. 

The network architecture itself, though physically separated into a process bus and a station bus, presents distinct vulnerabilities. The station bus is particularly critical, as it relies on standard IT protocols such as Transmission Control Protocol (TCP)/Internet Protocol (IP) and Manufacturing Message Specification (MMS) for configuration, monitoring, and non-time-critical communication. These widely used protocols are well-understood by attackers and have a larger ecosystem of available malicious tools, making them a primary vector for intrusion. An attacker gaining access to the station bus could push malicious configuration files to vIEDs, subtly altering protection settings without triggering immediate alarms. Furthermore, the use of Software-Defined Networking (SDN) to manage these networks introduces the SDN controller as another high-value target and compromising it would grant an attacker control over the entire substation's communication fabric.

Consequently, a robust defense-in-depth security architecture is not merely recommended but essential. This requires more than just perimeter defense. It mandates strict micro-segmentation to control traffic between vIEDs, strong authentication and access controls for all management interfaces, continuous monitoring of network traffic for anomalies, and sophisticated intrusion detection systems tailored for IEC 61850 protocols~\cite{Gaspar2023}. Finally, redundancy becomes crucial not only at the communication level, through protocols like Parallel Redundancy Protocol (PRP) and High-availability Seamless Redundancy (HSR), but also at the virtualization layer, where redundant virtual instances ensure failover capabilities in the event of a cyber-attack or a software failure~\cite{Tobar-Rosero2025, Gaspar2023}.

In summary, vIEDs represent a transformative step in substation automation, offering scalable, flexible, and cost-effective alternatives to traditional hardware-based devices. However, this transformation comes with trade-offs. The deterministic performance and physical isolation of conventional IEDs are replaced by systems whose reliability hinges on the robustness of IT and networking infrastructure. Furthermore, the shift to centralized, software-defined architectures increases exposure to cybersecurity threats, necessitating a holistic and proactive defense strategy. Therefore, rigorous evaluation of vIED performance, reliability, and security is essential prior to deployment. This paper aims to contribute to this field by presenting the design, implementation, and performance evaluation of such a \gls{vied}, demonstrating its viability for real-time applications in digital substations.
}

\section{The Virtual IED Architecture}\label{sec:architecture}

\begin{figure}[!ht]
  \centering
\includegraphics[width=1\columnwidth]{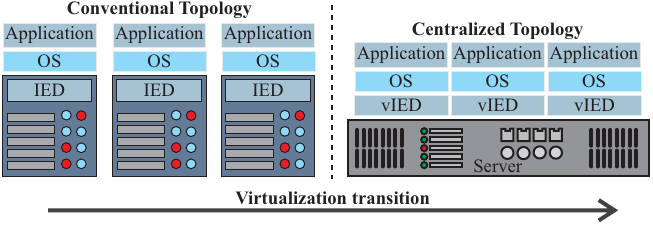}
  \caption{Transition from multiple standalone \glspl{ied} to virtualized IEDs.}\label{fig:vIED-Transition}
\end{figure}

As mentioned before, a \gls{vied} represents a software-based implementation of an IED, engineered to emulate the protection and control capabilities of traditional hardware IEDs within digital substations. This virtualization reflects a shift from hardware-dedicated systems to software-defined protection systems, as illustrated in Fig.~\ref{fig:vIED-Transition}.

The following sections provide an in-depth description of the implemented \gls{vied} architecture, focusing on its data processing components and protection algorithms.

\subsection{Overview of Algorithms}

\begin{figure}[!htb]
  \centering
\includegraphics[width=1\columnwidth]{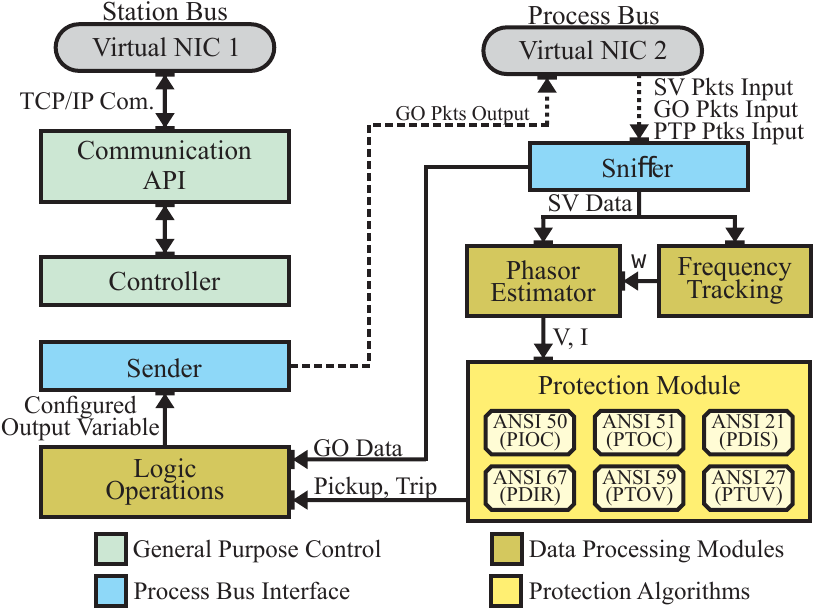}\caption{Software architecture of the vIED divided into modules.} 
  \label{fig:vIED-Architecture}
\end{figure}

The architecture of \gls{vied}, illustrated in Fig.~\ref{fig:vIED-Architecture}, consists of several modules, including general purpose control, the communication interface with the process bus, data processing, and protection algorithms.

\subsubsection{General-purpose control module} 
This module oversees \gls{vied} operations and handles communication with the station bus through TCP/IP. It also provides a system configuration interface and real-time monitoring of virtual equipment. 

\subsubsection{Communication Interface Module}

{\color{black} Acting as the interface between the \gls{vied} and the process bus, this module is responsible for subscribing to and processing incoming IEC 61850 messages. Specifically, it decodes \gls{sv} streams to extract raw current and voltage samples for subsequent processing by the data acquisition module and parses \gls{goose} messages to update the status of external binary signals. In addition, it handles the publication of outgoing \gls{goose} messages, such as trip or block commands, to the communication network.

In order to ensure high-performance operation with minimal latency, the communication stack was entirely implemented, using a custom protocol handler developed at the application level. This implementation leverages a low-level interface with the NIC driver, bypassing the conventional kernel network stack.}

\subsubsection{Data processing module}
This unit is essential for signal conditioning and measurement extraction. It processes incoming \gls{sv} messages to derive the current and voltage values required for phasor estimation, while \gls{goose} messages serve as digital input.

\subsubsection{Protection algorithms module} 
This component implements primary protection functions, such as overcurrent, distance, overvoltage, undervoltage, and directional protection schemes. These algorithms process data to identify faults, triggering trip and pickup signals to ensure prompt protective actions in abnormal operating conditions. Detailed descriptions of specific algorithms are provided in the section~\ref{sec:protection}.

\subsection{Frequency Tracking Algorithm}

\begin{figure}[!ht]
  \centering
  \includegraphics[width=1\columnwidth]{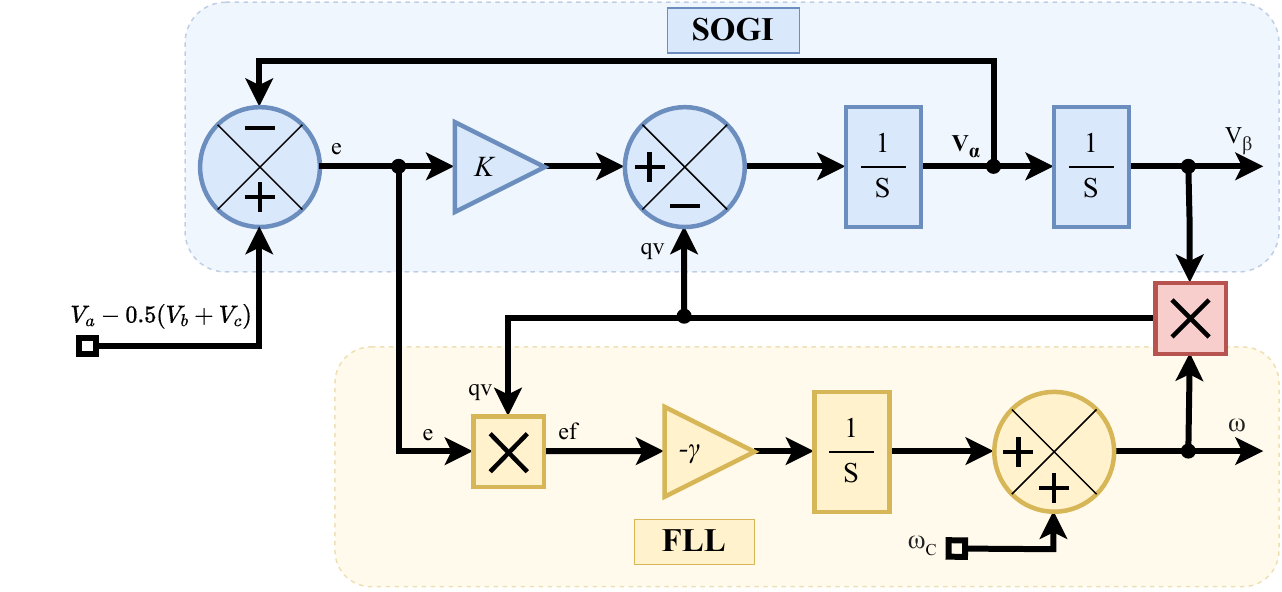}
  \caption{Adapted SOGI FLL algorithm for frequency tracking.}\label{fig:FrequencyTrackingAlgorithm}
\end{figure}

The frequency tracking function is essential for accurate phasor estimation, as it enables the adaptation of protection algorithms to dynamic power system conditions. The implemented frequency tracking algorithm continuously monitors the system frequency and dynamically adjusts the phasor estimation process to compensate for frequency deviations, thereby maintaining stability in real-time applications. The algorithm, illustrated in Fig.~\ref{fig:FrequencyTrackingAlgorithm}, uses a frequency-locked loop (FLL) of the second-order generalized integrator (SOGI) based on the methodology presented in~\cite{Bamigbade2022}.

Additionally, frequency tracking is restricted within a predefined range between 40 Hz and 70 Hz. If the system frequency exceeds these limits, it is clamped to the nearest boundary, thereby preventing inaccuracies in the phasor estimation process due to extreme frequency deviations. 

\subsection{Phasor Estimation}

\begin{figure}[!ht]
  \centering
  \includegraphics[width=1\columnwidth]{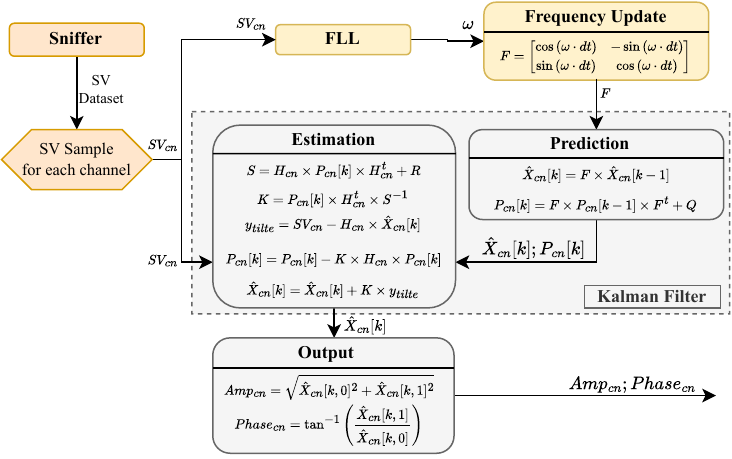}
  \caption{Nonlinear Kalman filtering algorithm used for phasor estimation.}\label{fig:PhasorEstimationAlgorithm}
\end{figure}

\textcolor{black}{The phasor estimation algorithm is responsible for accurately determining the magnitude and phase angle of current and voltage signals acquired from the process bus. These phasor quantities serve as critical inputs to various protection functions. To ensure robustness under dynamic operating conditions, the algorithm employs a nonlinear Kalman filtering approach, which enhances estimation accuracy by effectively suppressing measurement noise and adapting to signal variability. The overall structure of the proposed algorithm is illustrated in Fig.~\ref{fig:PhasorEstimationAlgorithm}, and is based on the methodology presented in~\cite{Khodaparast2022}.}

\textcolor{black}{The nonlinear Kalman filter was selected due to its optimal estimation capabilities in environments characterized by measurement noise, as well as its ability to reliably track phasor dynamics during transient disturbances. Nevertheless, alternative phasor estimation techniques, such as cosine filters, can also be implemented within the \gls{vied} framework, depending on the application requirements and computational constraints.}

\subsection{Protection Schemes}
\label{sec:protection}

The \gls{vied} implements multiple protection functions to assess its performance within a virtualized environment. The following topics describe the protection schemes implemented, including overcurrent, distance, overvoltage, undervoltage, and directional protection.

\subsubsection{Instantaneous and Time Overcurrent Protection (PIOC and PTOC)} 

These protection functions, also referred to as ANSI 50 and 51, are among the most commonly used protection mechanisms in power systems. They continuously monitor the current within a protected zone and initiate a timer when the current exceeds predefined threshold values, known as the pickup current.

\begin{figure}[!ht]
  \centering
  \includegraphics[width=1\columnwidth]{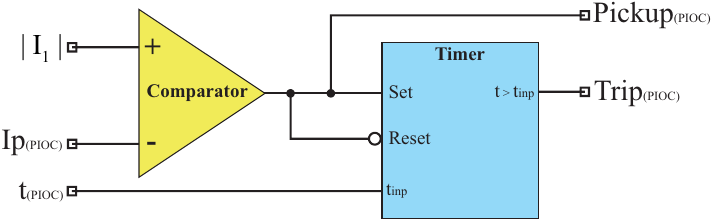}
  \caption{PIOC operation principle with pickup and trip output signals.} \label{fig:PIOCOperation}
\end{figure}

The PIOC can operate immediately or with a fixed delay, as configured by the operator. In instantaneous mode, it responds immediately upon the current exceeding the pickup value. Alternatively, if configured with a fixed delay, the PIOC waits for the specified time before initiating an action, allowing it to coordinate with other protection devices. Fig. \ref{fig:PIOCOperation} illustrates the principle of PIOC operation.

\begin{figure}[!ht]
  \centering
  \includegraphics[width=1\columnwidth]{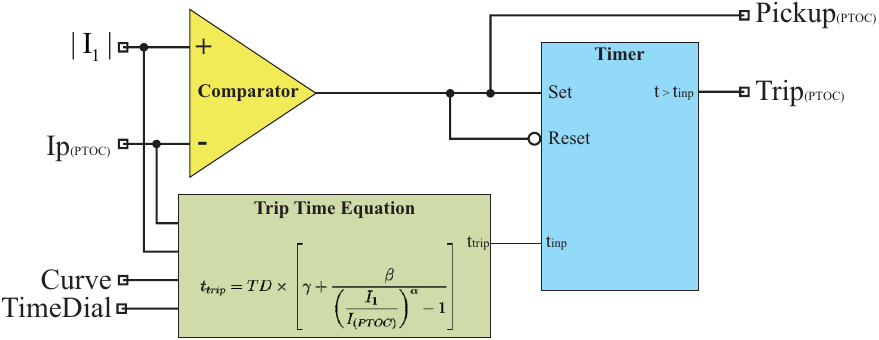}
  \caption{PTOC operation principle with pickup and trip output signals.} \label{fig:PTOCOperation}
\end{figure}

In contrast, the PTOC function incorporates an inverse time-delay characteristic, which is particularly effective for identifying prolonged overload conditions. Unlike the PIOC, which provides immediate or fixed-time protection, the PTOC's response time is inversely related to the magnitude of the overcurrent, i.e., the higher the current, the shorter the delay. This inverse time characteristic facilitates coordination among protection devices at various system levels. Fig. \ref{fig:PTOCOperation} illustrates the operating principle of the PTOC, highlighting the time-delay response based on fault severity.

\subsubsection{Distance Protection (PDIS)} 

The PDIS represents the ANSI 21 function and operates by assessing the apparent impedance, determining whether it falls within a predefined protection zone corresponding to the line impedance that the relay is monitoring. If the impedance is within this zone, a timer is activated, and a trip signal is issued if the fault is not cleared within the operator-defined time threshold. Additionally, PDIS provides four distinct zone types from which the user can choose. These include impedance, admittance, reactance, and quadrilateral zones, each offering unique characteristics for fault detection and discrimination.

\begin{figure}[!ht]
  \centering
  \includegraphics[width=1\columnwidth]{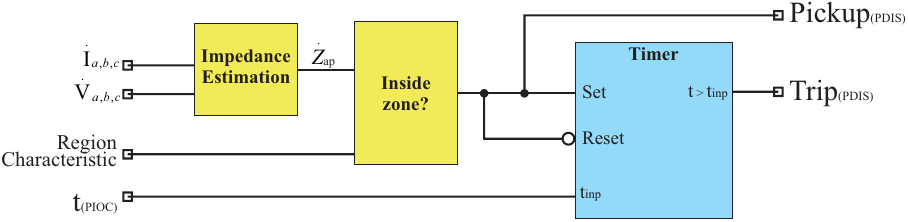}
  \caption{PDIS operation principle with pickup and trip output signals.}\label{fig:DistanceProtectionOperation}
\end{figure}

The PDIS operation scheme is depicted in Fig. \ref{fig:DistanceProtectionOperation}, highlighting the relay's response to fault conditions based on the measured impedance.

\subsubsection{Directional Protection (PDIR)} 

\begin{figure}[!ht]
  \centering
  \includegraphics[width=1\columnwidth]{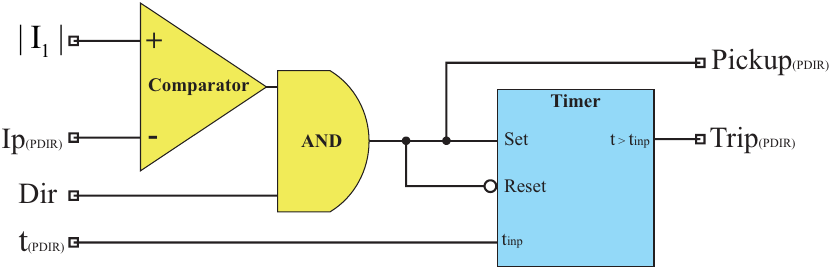}
  \caption{PDIR operation principle with pickup and trip output signals.}\label{fig:DirectionalProtectionOperation}
\end{figure}

This protection, refered by ANSI 67, operates on a principle similar to that of PIOC, incorporating an additional logic condition to verify the direction of the current flow by using an AND gate to the protection logic, as shown in Fig. \ref{fig:DirectionalProtectionOperation}.

The current direction is determined using the quadrature polarization method, which classifies the current as forward or reverse based on its angle relative to a reference quantity. This directional logic enhances fault discrimination by ensuring the protection responds only to faults within the designated direction of coverage.

% The current direction is determined using the quadrature polarization method, with forward and reverse regions illustrated for phase A in Fig. \ref{fig:DirectionalProtectionRegion}. The current flows in the forward direction when its angle lies within the blue region and in the reverse direction otherwise.

% \begin{figure}[!ht]
%   \centering
%   \includegraphics[width=0.4\columnwidth]{pdir-region.pdf}
%   \caption{PDIR Operation Regions -- forward and reverse delimitation for the phase A current direction.}
%   \label{fig:DirectionalProtectionRegion}
% \end{figure}

\subsubsection{Overvoltage and Undervoltage (PTOV and PTUV)} 

These functions, defined by ANSI 59 and 27, respectively, are designed to monitor voltage levels and trigger trip signals when the measured voltage deviates from predefined safe operating thresholds. 

Both functions operate on the same fundamental principle: they continuously compare the measured voltage against a user-defined threshold. When a threshold violation is detected, a timer is initiated, and if the condition persists beyond the configured delay, a trip signal is issued to isolate the faulted section. Fig.~\ref{fig:PTOVOperation} illustrates the operating logic of the PTOV function, where a trip is triggered when the measured voltage rises above the threshold. The PTUV function follows a similar logic but in the opposite direction, i.e., initiating a trip when the voltage falls below the set threshold.

\begin{figure}[!ht]
  \centering
  \includegraphics[width=1\columnwidth]{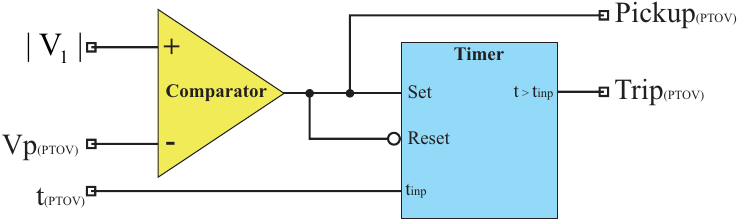}
  \caption{PTOV operation principle with pickup and trip output signals.}\label{fig:PTOVOperation}
\end{figure}

\begin{figure*}[!b]
  \centering
  \setcounter{figure}{10}
  \includegraphics[width=0.8\linewidth]{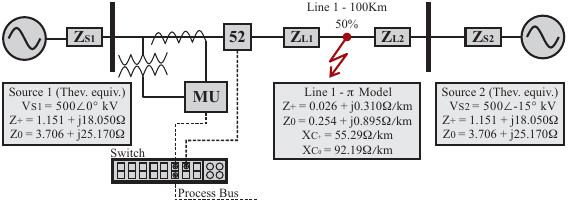}
  \caption{Simulated 500 kV transmission line model.}
  \label{fig:TestSystem}
\end{figure*}

\subsection{Server-based Virtual IED Deployment}
The \gls{vied} was implemented on a server that utilizes \glspl{vm} running independent instances of a Linux operating system. This VM-based approach enables each vIED to function as a fully isolated software-based protection device within the substation network. To achieve high-speed processing and deterministic behavior, critical for protection systems, the VM codebase was developed in low-level programming languages such as C/C++, ensuring strict adherence to timing requirements in data handling and algorithm execution.

In particular, VMs do not include a graphical user interface. Instead, all configuration and monitoring tasks were managed externally through a specialized desktop application developed for the Windows operating system. This software serves as the primary user interface, providing a centralized platform for streamlined configuration, real-time monitoring, and adjustment of the vIED's operational parameters.

This deployment method is particularly suited for commercial distribution, as the vIED can be packaged and provided as a compressed VM image file. This file can then be easily deployed on a suitable virtualization platform, allowing the utilities to integrate and configure the vIED within their existing digital substation infrastructure, thereby expanding the flexibility and scalability of protection system implementations.

\section{Test Setup}
\label{sec:test}

Testing was carried out in a virtual environment, employing a virtual Test Set designed to assess the performance of the \gls{vied} under different fault conditions. To ensure a precise emulation of the \gls{vied} protection system behavior, a transmission line model was created in PSCAD. The simulated waveforms were then exported to the virtual Test Set, which reconstituted them as test signals, simulating fault events and recording the \gls{vied} response times for each scenario. An overview of this test setup is shown in Fig. \ref{fig:TestSetup}.

\begin{figure}[H]
  \centering
  \addtocounter{figure}{-2}
  \includegraphics[width=1\columnwidth]{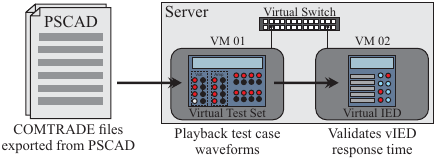}
  \caption{Test setup used to evaluate the \gls{vied} performance.}
  \label{fig:TestSetup}
\end{figure}
\addtocounter{figure}{1}
\subsection{Simulated Transmission Line}

A 500-kV transmission line, 100 km in length, was analyzed with grid equivalents operating at 60 Hz connected at both terminals. The line segment was represented using a single $\pi$ section model, as illustrated in Fig. \ref{fig:TestSystem}.

Fault scenarios were introduced at the midpoint of the transmission line, located 50 km from each terminal. A comprehensive range of fault conditions was covered, simulating four different fault types (phase-to-ground, phase-to-phase, phase-to-phase-to-ground and three-phase). Each fault scenario included resistances of 0, 15, 30, and 50~$\Omega$, with fault inception angles of 0°, 45°, and 90°. This setup yielded a comprehensive set of 48 distinct fault conditions, providing a robust basis for evaluating the operational accuracy and response times of \gls{vied} across various fault conditions.

\subsection{Server Configuration}

The server was configured with Ubuntu Server 22.04 LTS, using a real-time kernel to meet the stringent timing requirements of the digital substation environment. Virtualization was achieved using KVM technology, with Open vSwitch (OVS) used for efficient virtual network management. Both \glspl{vm} (the test set and \gls{vied}) were assigned identical hardware resources, each provisioned with 2 CPU cores, 4 GB of RAM and 3 virtual network interfaces.

The network architecture consisted of three virtual switches, one for the station bus and the other two for the process bus. The station bus switch was connected to the physical network interface of the server and to one of the virtual interfaces of both the Test Set and \gls{vied}. This setup enabled TCP/IP communication between the configuration desktop software and the virtualized devices.

For the process bus, two dedicated virtual switches were used to implement the Parallel Redundancy Protocol (PRP), ensuring network redundancy for the \gls{sv} and \gls{goose} protocol messages. Each \gls{vm} was configured with dual virtual ports, each connected to a separate process bus switch. This configuration enabled seamless failover and uninterrupted data transmission on the process bus, ensuring reliable SV communication with PRP redundancy.

\subsection{Virtual IED Parameterization}

To validate the performance of \gls{vied}, four functions were selected according to the fault simulation conditions. The equipment was parameterized to respond to these simulated faults with the settings detailed in Table \ref{tab:ProtectionFunctionParametrization}.

\begin{table}[h!]
  \centering
  \caption{Virtual IED protection function parameterization.}\label{tab:ProtectionFunctionParametrization}
\renewcommand{\arraystretch}{1.3}
  \begin{tabular}{lll}
  \hline
  \textbf{Protection Function} & \textbf{Parameter} & \textbf{Value} \\ \hline
  \multirow{2}{*}{PIOC} & Pickup current & 2500 A \\ \cline{2-3} 
                        & Time delay      & 0 seconds \\ \hline
  \multirow{3}{*}{PTOC} & Pickup current  & 1300 A \\ \cline{2-3} 
                        & Curve           & US moderately inverse (U1) \\ \cline{2-3} 
                        & Time dial       & 1 \\ \hline
  \multirow{3}{*}{PDIS} & Reach           & 100\% of line impedance \\ \cline{2-3} 
                        & Characteristic  & MHO \\ \cline{2-3} 
                        & Time delay      & 0 seconds \\ \hline
  \multirow{2}{*}{PTUV} & Pickup voltage  & 0.9 pu \\ \cline{2-3} 
                        & Time delay      & 0.1 seconds \\ \hline
  \end{tabular}
  \end{table}

\section{Assessment of the Virtual IED Performance}
\label{sec:performance}

The vIED's performance was analyzed across 48 distinct fault scenarios to assess its response time and latency. Each fault scenario was repeated 50 times, with the response times for each fault condition recorded by the virtual Test Set. This extensive testing enabled a comprehensive assessment of the vIED's performance under various conditions, offering valuable insights into its response consistency and variability.

\begin{figure}[!ht]
  \centering
  \includegraphics[width=1\columnwidth]{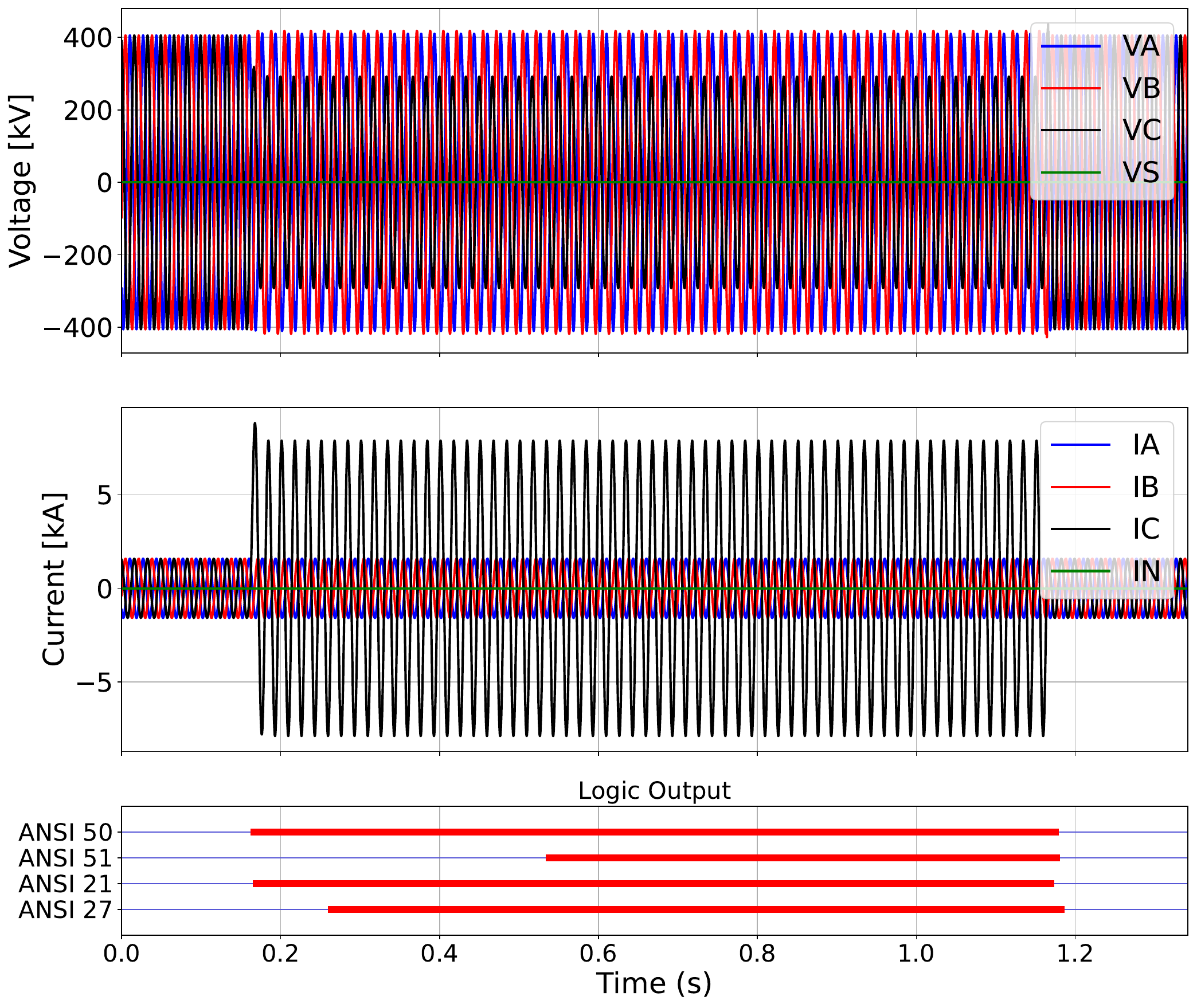}
  \caption{Response to a phase-to-ground fault with 15~$\Omega$ fault resistance (ANSI 50, 51, 21, 27).}\label{fig:oscillograph}
\end{figure}

Fig.~\ref{fig:oscillograph} presents an oscillograph of the \gls{vied}'s response to a phase-to-ground fault with a fault resistance of 15~$\Omega$. This scenario provides a clear illustration of the device's protection function sensibilization.

The response times for all fault conditions are summarized in Tables~\ref{tab:res_PIOC} to~\ref{tab:res_PTUV}. Each table presents the minimum, mean, and maximum response times, as well as the standard deviation for different fault resistances, offering a comprehensive view of the latency and stability of \gls{vied} under various fault conditions.

Table \ref{tab:res_PIOC} shows the response times of the PIOC function, which reveal a progressive increase in mean response time with higher resistance faults. At zero resistance, the mean response time is low and consistent at 3.7117 ms, while it rises to 9.0385 ms at 50 $\Omega$, indicating a predictable increase in latency. The standard deviation of 0.8726~ms at the highest resistance reflects moderate variability.

\begin{table}[!ht]\centering
  \small  
  \caption{\color{black} Timing latency of the PIOC protection function in the vIED.}
  \label{tab:res_PIOC}
  \renewcommand{\arraystretch}{1.3}
  \resizebox{\columnwidth}{!}{%
  \begin{tabular}{cccccc}
    \hline
  \begin{tabular}[c]{@{}c@{}}\textbf{Fault} \\
  	\textbf{Resistance}\end{tabular} & \textbf{Minimum} & \textbf{Mean} & \textbf{Maximum} & \begin{tabular}[c]{@{}c@{}}\textbf{Std.} \\
  	\textbf{Deviation}\end{tabular} \\
    \hline
    0 $\Omega$ & 2.9867 ms & 3.7117 ms & 5.0000 ms & 0.8462 ms \\
    15 $\Omega$ & 3.4367 ms & 4.3382 ms & 5.6200 ms & 0.7616 ms \\
    30 $\Omega$ & 4.2167 ms & 6.2318 ms & 8.5433 ms & 1.6483 ms \\
    50 $\Omega$ & 7.7700 ms & 9.0385 ms & 10.0033 ms & 0.8726 ms \\
    \hline
    \end{tabular}}
\end{table}

The response times of the PTOC function, as shown in Table \ref{tab:res_PTOC}, range from a minimum of 0.8605 ms to a maximum of 13.5905 ms. In contrast to the PIOC function, the PTOC response time remains largely unaffected by increasing fault resistance, with an average response time consistently around 2 ms. The low standard deviation across all fault resistances indicates stable performance, likely due to this function's inherent time delay, which minimizes the impact of transient fluctuations on response time.

\begin{table}[!ht]\centering
  \small 
  \caption{\color{black} Timing latency of the PTOC protection function in the vIED.}
  \label{tab:res_PTOC}
  \renewcommand{\arraystretch}{1.3}
  \resizebox{\columnwidth}{!}{%
  \begin{tabular}{cccccc}
    \hline
  \begin{tabular}[c]{@{}c@{}}\textbf{Fault} \\
  	\textbf{Resistance}\end{tabular} & \textbf{Minimum} & \textbf{Mean} & \textbf{Maximum} & \begin{tabular}[c]{@{}c@{}}\textbf{Std.} \\
  	\textbf{Deviation}\end{tabular} \\
    \hline
    0 $\Omega$ & 0.8605 ms & 1.7178 ms & 13.5905 ms & 1.1912 ms \\
    15 $\Omega$ & 0.8350 ms & 1.7428 ms & 2.9916 ms & 0.7471 ms \\
    30 $\Omega$ & 0.4174 ms & 1.9688 ms & 8.9574 ms & 0.9552 ms \\
    50 $\Omega$ & 1.0256 ms & 1.9687 ms & 9.5989 ms & 1.2172 ms \\
    \hline
    \end{tabular}}
  \end{table}

Table \ref{tab:res_PDIS} presents the response of the PDIS function, demonstrating a significant influence of the fault resistance on response time. The mean response time increases from 6.7757 ms at 0 $\Omega$ to 14.5585 ms at 30 $\Omega$, indicating a clear sensitivity to increasing resistance. Despite this notable trend, the low standard deviation of 1.1653 ms reflects relatively consistent performance across trials. This increase in response time with fault resistance underscores the sensitivity of the PDIS function to resistance changes, likely attributed to the phasor estimation process during the transient period.

\begin{table}[!ht]\centering
  \small 
  \caption{\color{black}  Timing latency of the PDIS protection function in the vIED.}
  \label{tab:res_PDIS}
  \renewcommand{\arraystretch}{1.3}
  \resizebox{\columnwidth}{!}{
  \begin{tabular}{cccccc}
    \hline
  \begin{tabular}[c]{@{}c@{}}\textbf{Fault} \\
  	\textbf{Resistance}\end{tabular} & \textbf{Minimum} & \textbf{Mean} & \textbf{Maximum} & \begin{tabular}[c]{@{}c@{}}\textbf{Std.} \\
  	\textbf{Deviation}\end{tabular} \\
    \hline
    0 $\Omega$ & 5.0367 ms & 6.7757 ms & 10.3633 ms & 1.6358 ms \\
    15 $\Omega$ & 6.1667 ms & 8.3268 ms & 11.2100 ms & 1.3700 ms \\
    30 $\Omega$ & 7.3033 ms & 14.5585 ms & 18.9533 ms & 1.1653 ms \\
    \hline
    \end{tabular}}
  \end{table}

Table \ref{tab:res_PTUV} summarizes the response times of the PTUV function, which exhibit moderate fault resistance sensitivity, with values ranging from 0.5133 ms to 9.5500 ms. The mean response time remains relatively low, even at higher resistances, indicating a rapid response. However, at 50 $\Omega$, the standard deviation increases to 3.8856~ms, higher than the other protection functions, suggesting some variability. This increased variation at 50 $\Omega$ can be attributed to a minimal voltage drop under high-resistance fault conditions, which can affect response time consistency.

\begin{table}[!ht]\centering
  \small 
  \caption{\color{black}Timing latency of the PTUV protection function in the vIED.}\label{tab:res_PTUV}
  \renewcommand{\arraystretch}{1.3}
  \resizebox{\columnwidth}{!}{
  \begin{tabular}{cccccc}
    \hline
  \begin{tabular}[c]{@{}c@{}}\textbf{Fault} \\
  	\textbf{Resistance}\end{tabular} & \textbf{Minimum} & \textbf{Mean} & \textbf{Maximum} & \begin{tabular}[c]{@{}c@{}}\textbf{Std.} \\
  	\textbf{Deviation}\end{tabular} \\
    \hline
    0 $\Omega$ & 0.5133 ms & 1.5098 ms & 7.6967 ms & 1.0802 ms \\
    15 $\Omega$ & 0.5233 ms & 1.6780 ms & 7.2867 ms & 1.2429 ms \\
    30 $\Omega$ & 0.5233 ms & 1.6422 ms & 6.8367 ms & 1.2948 ms \\
    50 $\Omega$ & 0.5133 ms & 3.6743 ms & 9.5500 ms & 3.8856 ms \\
    \hline
    \end{tabular}}
\end{table}

In general, the vIED performed reliably, with response times acceptable for virtualized protection systems. The variability was low except where the fault resistance affected certain functions, probably due to a delayed estimation of the phasor under dynamic fault conditions. Across all functions, the global mean response time was 5.5278 ms, with a standard deviation of 1.1332 ms, indicating stable performance in a virtual environment. These findings suggest that \gls{vied} could serve as a feasible alternative to physical devices for certain power system relaying applications.

\section{Conclusions}
\label{sec:conclusions}

This paper presents a comprehensive implementation and performance evaluation of a vIED designed to fulfill the protection and control functions traditionally managed by physical IEDs in digital substations. The vIED was implemented in a virtualized, IEC 61850-compliant environment and rigorously tested across 48 fault scenarios (each repeated 50 times) to assess its response time and reliability.

The evaluation demonstrated that the vIED delivered consistent and reliable performance, remaining within the expected operating ranges for virtualized protection systems. Although response time variability remained generally low, certain protection functions exhibited delayed responses at higher fault resistances, largely due to inherent delays in phasor estimation under dynamic fault conditions. This responsiveness within acceptable limits in various fault scenarios highlights the potential of \gls{vied} as a practical alternative solution for applications requiring consistent time-sensitive responses.

These findings support the feasibility of using virtualized protection equipment in digital substations. It has been demonstrated that a vIED can meet the response time and reliability standards traditionally associated with physical devices, underscoring the suitability of vIEDs for enhancing the flexibility, scalability, and cost-effectiveness of power grid protection systems.

Although these results confirm the overall feasibility, further research is essential before practical deployment, particularly regarding cybersecurity, redundancy, interoperability, network performance under high background load, and host-level failover behavior.

\bibliographystyle{IEEEtran} 
\bibliography{references.bib}

\vfill

\end{document}